\documentclass[12pt,a4paper]{article}
\usepackage{mathrsfs}
\usepackage{epsfig}
\begin{document}
\title{Pair production of the T-odd leptons at the $LHC$}
\author{Chong-Xing Yue, Yong-Zhi Wang, Wei Lui, Wei Ma  \\
{\small Department of Physics, Liaoning  Normal University, Dalian
116029, P. R. China}
\thanks{E-mail:cxyue@lnnu.edu.cn}}
\date{\today}

\maketitle
\begin{abstract}

\vspace{1cm}

The T-odd leptons predicted by the littlest $Higgs$ model with
T-parity can be pair produced via the subprocesses $gg\rightarrow
\ell^{+}_{H}\ell^{-}_{H}$, $q\overline{q}\rightarrow
\ell^{+}_{H}\ell^{-}_{H}$, $\gamma\gamma\rightarrow
\ell^{+}_{H}\ell^{-}_{H}$ and $ VV \rightarrow
\ell^{+}_{H}\ell^{-}_{H}$ ($V$=$W$ or $Z$) at the $CERN$ Large
Hadron Collider $(LHC)$. We estimate the hadronic production cross
sections for all of these processes and give a simply phenomenology
analysis. We find that the cross sections for most of the above
processes are very small. However, the value of the cross section
for the  $Drell-Yan$ process $q\overline{q}\rightarrow
\ell^{+}_{H}\ell^{-}_{H}$ can reach $270fb$.

 \vspace{2.0cm}

\end{abstract}

\newpage
\noindent{\bf 1. Introduction}

The $CERN$ Large Hadron Collider $(LHC)$ will soon go into full
operation and provide proton-proton collisions at a center-of-mass
$(c. m.)$ energy of 14$TeV$, a factor of 7 higher than the Tevatron.
There are strong theoretical reasons to expect that the $LHC$ will
discover new physics beyond the standard model $(SM)$ up to $TeV$.
Many popular new physics models beyond the $SM$ predict the
existence of the new charged leptons which are denoted as $L^{\pm}$.
In general, as long as they are not too heavy, these new particles
can be pair produced via the gluon fusion process $gg\rightarrow
L^{+}L^{-}$, the $Drell-Yan$ process $q\overline{q}\rightarrow
L^{+}L^{-}$, the photon-induced process $\gamma\gamma\rightarrow
L^{+}L^{-}$, and the weak gauge boson fusion process $VV\rightarrow
L^{+}L^{-}$ ( $V$=$W$ or $Z$) at the $LHC$.

Studying production and decay of the new charged leptons at current
or in future high energy collider experiments is of special
interest. Any signal for such kind of new fermions in future high
energy experiments will play an important role in testing the $SM$
flavor structure and discovery of new physics beyond the $SM$. This
fact has lead to many works involving the new charged leptons at
$e^{+}e^{-}$ colliders [1], $ep$ colliders [2], and hadron colliders
[3,4,5].

The littlest $Higgs$ model with T-parity, which is called the $LHT$
model [6], is one of the attractive little $Higgs$ models. The $LHT$
model predicts the existence of the T-odd $SU(2)$ doublet fermions.
These new fermions can produce rich phenomenology at present or in
future high energy collider experiments [7,8,9,10,11]. In Ref.[12],
we have studied pair production of the T-odd leptons in an
international linear $e^{+}e^{-}$ collider ($ILC$). Our numerical
results show that, in wide range of the parameter space, they can be
copiously produced in pairs. The possible signatures of the T-odd
leptons might be detected in future $ILC$ experiments.

Since it is regarded as the cross section for pair production of the
T-odd leptons is generally small at the $LHC$, so far there are few
works involved their directly production at the $LHC$. However,
considering the $LHC$ will go into operation and it is possible to
completely study the possible signals of the T-odd leptons at the
$LHC$, it is need to carefully study pair production of the T-odd
leptons at the $LHC$. So, in this note, we will discuss various
possible production channels of the T-odd lepton pairs at the $LHC$
and compare their values of the production cross sections for
different production channels. In the next section, we will give our
numerical results. Our conclusion and a simply phenomenology
analysis are given in the last section.

\noindent{\bf 2. The numerical results }

The $LHT$ model [6] is based on an $SU(5)/SO(5)$ global symmetry
breaking pattern. A subgroup $[SU(2)\times
U(1)]_{1}\times[SU(2)\times U(1)]_{2}$ of the $SU(5)$ global
symmetry is gauged, and at the scale $f$ it is broken into the $SM$
electroweak symmetry $SU(2)_{L}\times U(1)_{Y}$. T-parity is an
automorphism that exchanges the $[SU(2)\times U(1)]_{1}$ and
$[SU(2)\times U(1)]_{2}$ gauge symmetries. To simultaneously
implement T-parity, one needs to double the $SM$ fermion doublet
spectrum [6,7]. The T-even combination is associated with the $SM$
$SU(2)_{L}$ doublet, while the T-odd combination is its T-parity
partner. The T-odd fermion sector consists of three generations of
the mirror quarks and leptons with vectorial couplings under
$SU(2)_{L}\times U(1)_{Y}$. Only T-odd leptons are related our
calculation, we denote them by
\begin{equation}
 \left(
 \begin{array}{c}\nu^{1}_{H}\\\ell^{2}_{H}\end{array}\right),
 \hspace{1.5cm}
 \left(
  \begin{array}{c}\nu^{2}_{H}\\\ell^{2}_{H}\end{array}\right),
 \hspace{1.5cm}
  \left(
  \begin{array}{c}\nu^{3}_{H}\\\ell^{3}_{H}\end{array}\right)
\end{equation}
with their masses satisfying to first order in $\nu/f$[10]
\begin{eqnarray}
 M^{1}_{\nu_{H}}=M^{1}_{\ell_{H}},\hspace{0.5cm}
 M^{2}_{\nu_{H}}=M^{2}_{\ell_{H}},\hspace{0.5cm}
 M^{3}_{\nu_{H}}=M^{3}_{\ell_{H}}.
\end{eqnarray}
Here $\nu=246GeV$ is the electroweak scale and $f$ is the scale
parameter of the gauge symmetry breaking of the $LHT$ model. There
are $M^{i}_{l}=\sqrt{2}k^{i}_{l}f$, in which $k^{i}_{l}$ are the
eigenvalues of the mass matrix $k$ and their values are generally
dependent on the lepton species $i$. The coupling expressions of the
T-odd lepton to other particles are given in Refs.[7,10].

It is well known that, at the $LHC$, the gluon distribution function
is considerably larger than that for the quark. So we first consider
the gluon-induced production of the T-odd lepton pairs. In order to
produce the lepton pairs from gluon fusion mechanism, we must have a
virtual quark triangular loop, which is connected to the lepton
pairs by a neutral boson ( photon $\gamma$, gauge boson $Z$, or
scalar $H$). However, the contributions of the photon vector
coupling and the $Z$ vector coupling vanish due to Furry's theorem
[3]. From Refs.[7,10], one can see that the $Z$ axial-vector
coupling to the T-odd leptons is equal to zero and the scalar $H$
can not couple to the T-odd quarks $T_{-}$ and $D_{H}$. Thus, the
T-odd lepton pair can only be produced via the $s$-channel $H$
exchange diagram, in which the quark triangular loops include the
top quark $t$, the up-type T-odd quark $U_{H}$, and the T-even
vectorlike quark $T_{+}$.

Using the relevant Feynman rules given in Refs.[7,10], we can easily
calculate the hadronic production cross section $\sigma_{g}$  of the
gluon fusion process $gg\rightarrow H\rightarrow
\ell^{+}_{H}\ell^{-}_{H}$ at the $LHC$, which depend on the T-odd
quark masses, the T-odd lepton masses, the mixing parameter $x_{L}$,
and the scale parameter $f$. The mass of the T-even vectorlike top
quark $T_{+}$ can be determined by the free parameters $x_{L}$ and
$f$. To simply our calculation, we will take the free parameter
$x_{L}=0.5$ and the $Higgs$ boson mass $M_H=120GeV$. For the T-odd
quark masses, we will take
$M^{1}_{U_{H}}=M^{2}_{U_{H}}=M^{3}_{U_{H}}=M_{U}$ and assume that
their values are smaller than $4TeV$. $M_{\ell}$ is the T-odd lepton
mass taken as
$M^{1}_{\ell_{H}}=M^{2}_{\ell_{H}}=M^{3}_{\ell_{H}}=M_{\ell}$ and
assumed that its value is in the range of $200GeV\sim 1000GeV$.
Recently, the improved parton distribution functions ($PDFs$) have
been given in the literature, for example in Refs.[13,14]. However,
as numerical estimation, we will use $CTEQ6L$ $PDF$[15] for the
gluon $PDF$ and assume that the renormalization scale $\mu_{R}$ and
the factorization scale $\mu_{F}$ have the relation
$\mu_{F}=\mu_{R}=2M_{\ell}$.   Our calculation results show that the
production cross section $\sigma_{g}$ is very small and its value is
smaller than $1fb$ in most of the parameter space of the $LHT$
model. This is because the coupling of the $Higgs$ boson $H$ to the
T-odd leptons is proportion to the factor $\nu^{2}/f^{2}$ and the
production cross section $\sigma_{g}$ is suppressed at least by the
factor $\nu^{4}/f^{4}$.

The T-odd lepton pairs can also be produced via the $Drell-Yan$
process and the $\gamma\gamma$ fusion mechanism. The former process
is induced by the $s$-channel $\gamma$ exchange and $Z$ exchange,
while the latter process can proceed via the $t$-channel and
$u$-channel T-odd lepton exchanges. Since the $H\gamma\gamma$ and
$H\ell^{+}_{H}\ell^{-}_{H}$ couplings are very small, we will ignore
the contributions of the $s$-channel process
$\gamma\gamma\rightarrow H \rightarrow \ell^{+}_{H}\ell^{-}_{H}$. In
this case, it is obvious that the production cross sections of the
subprocesses $q\overline{q}\rightarrow\ell^{+}_{H}\ell^{-}_{H}$ and
$\gamma\gamma\rightarrow\ell^{+}_{H}\ell^{-}_{H}$ are only dependent
on the model dependent parameter --- the T-odd lepton mass
$M_{\ell}$.

\begin{figure}[htb]
\begin{center}
\epsfig{file=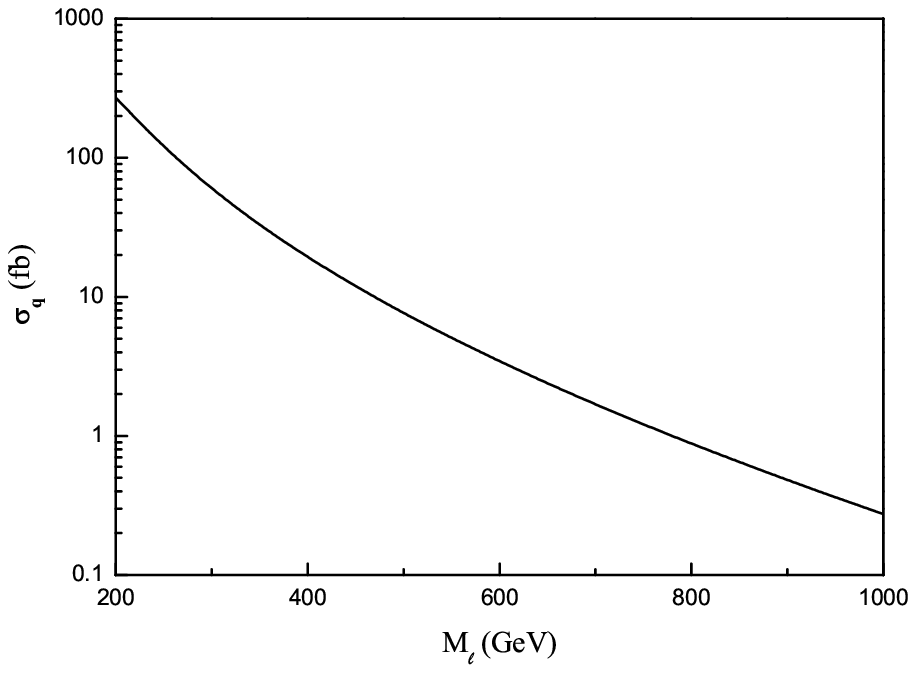,width=220pt,height=200pt} \put(-115,-10)
{(a)} \put(115,-10){(b)}
\epsfig{file=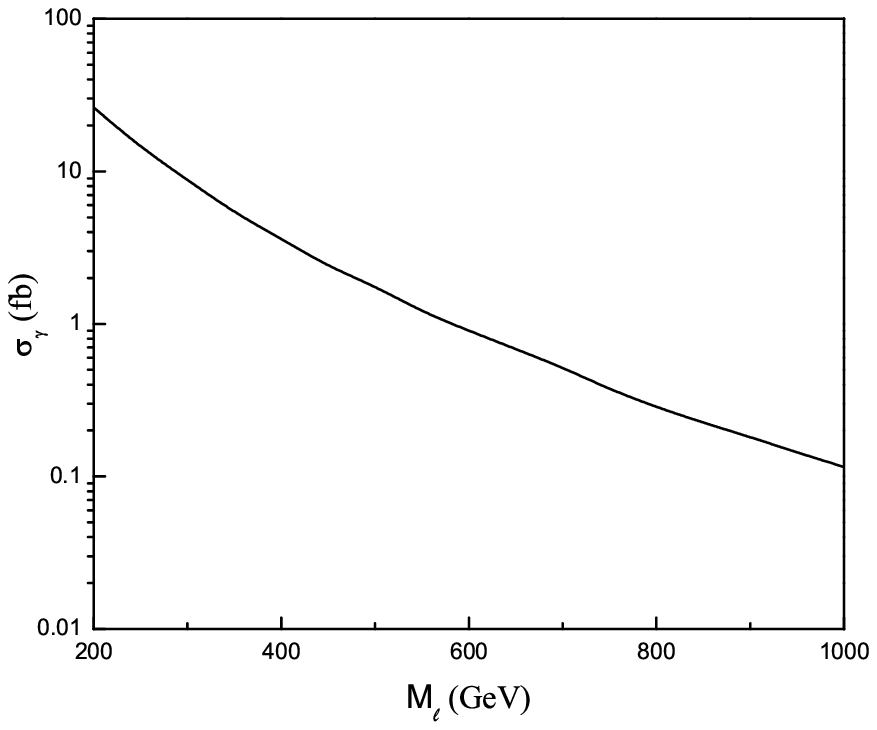,width=230pt,height=210pt} \caption{The
production cross sections $\sigma_{q}$ (a) and $\sigma_{\gamma}$
(b)as function of the  T-odd \hspace*{1.92cm}lepton mass
$M_{\ell}$.} \label{ee}
\end{center}
\end{figure}

Using the equivalent photon approximation$(EPA)$ approach [16], the
hadronic cross section $\sigma_{\gamma}$ at the $LHC$ can be obtain
by folding the cross section for the subprocess
$\gamma\gamma\rightarrow\ell^{+}_{H}\ell^{-}_{H}$ with the photon
distribution function $f_{\gamma/p}$, which can be written as
$f_{\gamma/p}=f_{\gamma/p}^{el}+f_{\gamma/p}^{inel}$.
$f_{\gamma/p}^{el}$ and $f_{\gamma/p}^{inel}$ are the elastic and
inelastic components of the equivalent photon distribution of the
proton, which have been extensively studied in Refs.[17, 18]. Our
numerical results are given in Fig.1, in which the pair production
cross sections $\sigma_{q}$(Fig.1a) and $\sigma_{\gamma}$(Fig1.b)
are plotted as functions of the T-odd mass $M_{\ell}$. In our
numerical calculation, we have assumed that the photon distribution
function $f_{\gamma/p}$ includes both the elastic and inelastic
components of the equivalent photon distribution of the proton
 and have taken the $CTEQ6L$ $PDFs$ for the quark
distribution functions $f_{q_{i}/p}$.  One can see from Fig.1 that
the production cross section of the T-odd lepton pairs induced by
the $Drell-Yan$ process is significantly larger than that for the
$\gamma\gamma$ fusion mechanism. For $200GeV\leq M_{\ell}\leq
1000GeV$, the values of the cross sections $\sigma_{q}$ and
$\sigma_{\gamma}$ are in the ranges of $270.6fb \sim 0.3fb$ and
$26.4fb \sim 0.11fb$, respectively. If we assume the yearly
integrated luminosity $\pounds_{int}$ = 100$fb^{-1}$ for the $LHC$
with the $c.m.$ energy $\sqrt{s}=14TeV$, then there will be several
up to ten thousands of the $\ell ^{+}_{H} \ell ^{-}_{H}$ events to
be generated per year.

It is well known that one of the main tasks of the $LHC$ is to
determine whether the breaking of the electroweak symmetry is due to
the $SM$ $Higgs$ boson. The vector boson fusion $(VBF)$ mechanism is
the second most copious source for the $SM$ $Higgs$ boson production
at the $LHC$, which can be used to directly study the exact dynamics
of the electroweak symmetry breaking. The new charged lepton pairs
can also be produced via the $VBF$ mechanism at the $LHC$ [4].  Pair
production of the T-odd leptons predicted by the $LHT$ model can be
induced by the $s$-channel processes $W^{\pm} W^{\mp}\rightarrow
\gamma, Z, H\rightarrow \ell ^{+}_{H} \ell ^{-}_{H}$ and
$ZZ\rightarrow H \rightarrow \ell^{+}_{H} \ell^{-}_{H}$, and can
also proceed via the $t$-channel T-odd neutrino $\nu_{H}$ exchange
and T-odd lepton $ \ell_{H}$ exchange for the processes $W^{\pm}
W^{\mp}\rightarrow \ell ^{+}_{H} \ell ^{-}_{H}$ and $ZZ\rightarrow
\ell^{+}_{H} \ell^{-}_{H}$, respectively.

\begin{figure}[htb]
\begin{center}
\epsfig{file=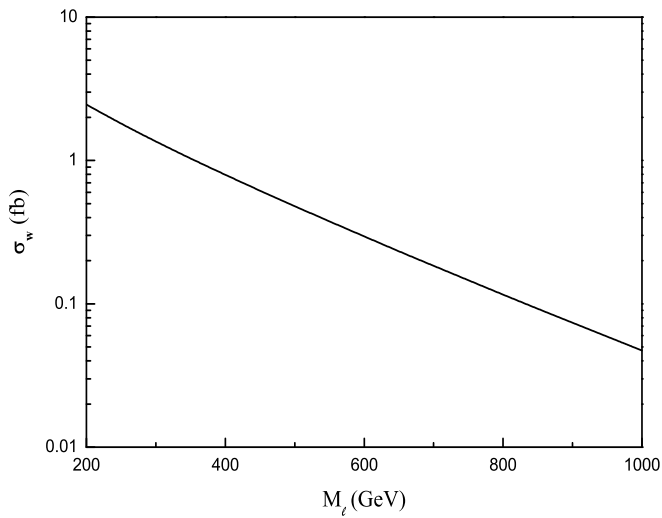,width=285pt,height=240pt} \caption{The
production cross section $\sigma_{W}$ for the subprocess
$W^{+}W^{-}\rightarrow\ell^{+}_{H}\ell^{-}_{H} $ as a
\hspace*{1.9cm} function of the T-odd lepton mass $M_{\ell}$.}
\label{ee}
\end{center}
\end{figure}

For $M_{\ell}>200GeV$, the $c. m. $ energy $ \sqrt{\hat{s}}$ of the
subprocesses $W^{\pm}W^{\mp} \rightarrow \ell^{+}_{H}\ell^{-}_{H}$
is larger than $400GeV $ and there is $  \hat{s}\gg M^{2}_{W}$. In
this energy region, the gauge bosons $W^{\pm}$ coming from quark or
antiquark can be treated as on-shell. So, in our estimating the
hadronic production cross section of the subprocesses
$W^{\pm}W^{\mp} \rightarrow \ell^{+}_{H}\ell^{-}_{H}$, we will adopt
the effective $W$-boson approximation $(EWA)$ method [19] and
include all contributions of the longitudinal and transverse W-boson
components. The structure functions for the gauge bosons $W^{\pm}$
are taken as the forms given by Ref.[20], in which authors have
shown that the $EWA$ method approximates very well the exact result.
Considering the coupling of the $Higgs$ boson $H$ to the T-odd
lepton pairs, which can be approximately written as $
-\frac{iM_{l}}{\nu}\frac{\nu^{2}}{4f^{2}}$, is very small, we have
neglected the contributions of $H$ exchange. Our numerical results
are shown in Fig.2. One can see from Fig.2 that the value of the
production cross section can only reach 2.5$fb$ in most of the
parameter space, which is smaller than that for the $\gamma\gamma$
fusion mechanism.

Similar with that for the processes $W^{\pm} W^{\mp}\rightarrow \ell
^{+}_{H} \ell ^{-}_{H}$, we can also neglect the contributions of
the $s$-channel $H$ exchange to the process $ZZ\rightarrow
\ell^{+}_{H}\ell^{-}_{H}$. Thus, this process is mainly induced by
the $t$-channel and $u$-channel T-odd lepton exchanges. It is well
known that the structure function for the gauge boson $Z$ is smaller
than that for the gauge boson $W$, the hadronic cross section for
the subprocess $ZZ\rightarrow \ell^{+}_{H}\ell^{-}_{H}$ is
suppressed at least by an order of magnitude compared to that for
the subprocess $W^{+}W^{-}\rightarrow \ell^{+}_{H}\ell^{-}_{H}$. For
$200GeV\leq M_{\ell}\leq 1000GeV$, its value is in the range of
$1.35\times 10^{-1}fb \sim 2.5\times 10^{-3}fb$. We further estimate
the hadronic cross section for the subprocess $Z\gamma\rightarrow
\ell^{+}_{H}\ell^{-}_{H}$, our numerical results show that its value
is in the range of $1.25fb \sim 6.9\times 10^{-3}fb$ for $200GeV\leq
M_{\ell}\leq 1000GeV$.

\noindent{\bf 3. Conclusions and discussions}

In this letter, we consider various possible production channels of
the T-odd lepton pairs at the $LHC$. We find that the most important
production channel for the T-odd lepton pairs predicted by the $LHT$
model is the $Drell-Yan$ process, i.e. $q\overline{q}\rightarrow
\ell^{+}_{H}\ell^{-}_{H}$. For 200$GeV\leq M_{\ell} \leq$ 1000$GeV$,
the value of the hadronic production cross section is in the range
of 270.6$fb$$\sim$ 0.3$fb$. The second important production channel
is the $\gamma\gamma$ fusion mechanism, i.e.
$\gamma\gamma\rightarrow\ell^{+}_{H}\ell^{-}_{H}$. With reasonable
values of the free parameters for the $LHT$ model, the production
cross section value can reach $26.4fb$.

The T-odd leptons are always heavier than the T-odd gauge boson
$B_{H}$. They are lighter or heavier than the T-odd gauge bosons
$W_{H}$ and $Z_{H}$, which depend on the values of the coupling
parameter $k_{l}= k^{1}_{l}= k^{2}_{l}= k^{3}_{l}$. For
$k_{l}<0.45$, there is $M_{l}< M_{Z_{H}}\approx M_{W_{H}}$. Thus,
 for $k_{l}>0.45$, the possible decay modes of the T-odd lepton $\ell_{H}$ are
$B_{H}l$, $Z_{H}l$, and $W_{H}\nu$. However, as long as $k_{l}\leq 1
$, $\ell_{H}$ mainly decays to $B_{H}\ell$ and there is $Br
(\ell_{H}\rightarrow B_{H}\ell) \simeq 100\%$ for $k_{l}<0.45$. If
we assume that the T-odd lepton $\ell_{H}$ decays to $B_{H}\ell$,
 then pair production process of the T-odd leptons at the $LHC$, i.e.
 the process $P P \rightarrow\ell^{+}_{H}\ell^{-}_{H}+X$, can give rise to
 the $\overline{B_{H}}B_{H}l^{+}l^{-}$ final state. The
new gauge boson $B_{H}$ predicted by the $LHT$ model is the stable
and lightest T-odd particle, which can be seen as an attractive dark
matter candidate [7]. Certainly, if the T-parity is violated by
anomalies, the T-odd gauge boson $B_{H}$ can decay into the $SM$
gauge boson pairs $WW$ and $ZZ$ [21,22]. If we assume that T-parity
is strictly conserved and the T-odd gauge boson $B_{H}$ can be seen
as missing energy, then pair production of the T-odd leptons at the
$LHC$ can induce the opposite-sign same-flavor leptons plus missing
energy ($\ell^{+}\ell^{-}$+$\not\!\!E$) signature. The main
backgrounds for this kind of signals come from the $SM$ processes
$PP\rightarrow$ $ZZ$ + $X$, $PP\rightarrow \tau\overline{\tau}$ +
$X$, and $PP\rightarrow W^{+}W^{-}$ + $X$, in which the lepton
$\tau$ and the gauge bosons $Z$ and $W$ decay leptonically. Detailed
analysis of the signals and backgrounds have been given in Ref.[23].
They have shown that the $ZZ$ background can be eliminated by
calculating the invariant mass of the charged lepton pair, which can
reconstruct the $Z$ boson mass. It is possible to extract the
$\ell^{+}\ell^{-}$+$\not\!\!E$ signal from the $\tau\overline{\tau}$
background by applying suitable cuts. The $W^{+}W^{-}$ background is
much large, so it is very challenging to extract the
$\ell^{+}\ell^{-}$+$\not\!\!E$ signal from this background. However,
it has been shown that the appropriate cuts can significantly
suppress the $W^{+}W^{-}$ background [24]. Certainly, detailed
confirmation of the observability of the
$\ell^{+}\ell^{-}$+$\not\!\!E$ signal, which comes from the
subprocess $q\overline{q}\rightarrow \ell^{+}_{H}\ell^{-}_{H}$,
would require Monte-Carlo simulations of the signal and background.

Although the contributions of the $\gamma\gamma$ fusion mechanism to
pair production of the T-odd lepton pairs at the $LHC$ are much
smaller than those for the $Drell-Yan$ process. For $M_{\ell} \geq$
200$GeV$, its production cross section can only reach $26.4fb$.
However, due to absence of the proton remants, this production
process has clean experimental conditions and well defined final
states, which can be selected and precisely reconstructed [25]. The
$SM$ backgrounds coming from the partonic interactions, such as
$PP\rightarrow W^{+}W^{-}$, $ZZ$ , and $\tau\overline{\tau}$, can be
significantly omitted by using the large rapidity gap technique and
the dedicated very forward detectors [26]. The irreducible
backgrounds, such as $\gamma\gamma\rightarrow W^{+}W^{-}$ and
$\gamma\gamma\rightarrow \tau\overline{\tau}$ can be largely
suppressed by applying acceptance cuts. Certainly, detailed detector
simulation studies are needed.

\vspace{1.0cm}

\noindent{\bf Acknowledgments}

This work was supported in part by the National Natural Science
Foundation of China under Grants No. 10675057, Specialized Research
Fund for the Doctoral Program of Higher Education (SRFDP) (No.
200801650002), the Natural Science Foundation of the Liaoning
Scientific Committee(No. 20082148), and Foundation of Liaoning
Educational Committee(No. 2007T086).

\end{document}